\documentclass[sigconf]{acmart}

\usepackage{url}
\usepackage{verbatim}
\usepackage{graphicx}
\usepackage{color}
\usepackage{tikz}
\usetikzlibrary{arrows,positioning,shapes,fit,calc}
\usepackage{smartdiagram}
\usepackage{algorithmic,algorithm}

\setcopyright{none}
\renewcommand\footnotetextcopyrightpermission[1]{} 
\AtBeginDocument{%
  }

\begin{document}

\title{Greybox fuzzing time-intensive programs}

%
\author{Steve Huntsman}
\email{steve.huntsman@str.us}
\orcid{0000-0002-9168-2216}


\begin{abstract}
We examine (directed) greybox fuzzing from a geometrical perspective, viewing dissimilarities on inputs and on control flow graphs (with dynamical statistics) as primitive objects of interest. We prototype and evaluate \texttt{GoExploreFuzz}, a greybox fuzzer for time-intensive programs that incorporates this perspective. The results indicate useful capabilities for greybox fuzzing that have hitherto been underutilized, notably quantifying the diversity of paths and autonomously tuning the ``bandwidth'' of mutations.
\end{abstract}

%


\maketitle

\section{\label{sec:Introduction}Introduction}

\begin{figure}[h]
  \centering
	\begin{tikzpicture}[->,>=stealth',shorten >=1pt,scale=0.4]
		\node [minimum height=.9cm, minimum width=1.8cm, draw, align=center] (v1) at (0,0) {promising \\ input};
		\node [minimum height=.9cm, minimum width=1.8cm, draw, align=center] (v2) at (8,0) {mutator};
		\node [minimum height=.9cm, minimum width=1.8cm, draw, align=center] (v3) at (16,0) {nearby \\ inputs};
		\node [minimum height=.9cm, minimum width=1.8cm, draw, align=center] (v4) at (16,-4) {target \\ program};
		\node [minimum height=.9cm, minimum width=1.8cm, draw, align=center] (v5) at (8,-4) {paths \\ in CFG};
		\node [minimum height=.9cm, minimum width=1.8cm, draw, align=center] (v6) at (0,-4) {curator};
		\path[->] (v6) edge node [left] {go} (v1);
		\path[->] (v2) edge node [above] {explore} (v3);
		\path[->] (v4) edge node [above] {map} (v5);
		\foreach \from/\to in {
			v1/v2, v3/v4, v5/v6}
			\draw (\from) to (\to);
	\end{tikzpicture}  
\caption{Greybox fuzzers curate and {\bf go} to promising inputs; they then {\bf explore} a space of ``nearby'' inputs by passing mutated variants to an instrumented target program to map paths in the CFG that inform subsequent iterations. 
  }
  \label{fig:greyboxFuzzingArchitecture}
\end{figure}
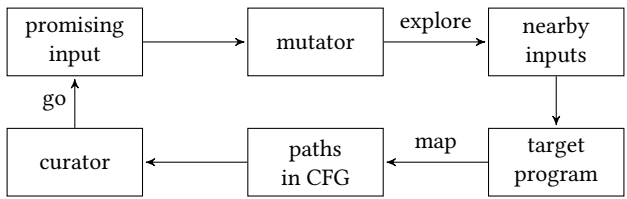

Greybox fuzzing \cite{bohme2016coverage,bohme2017directed,manes2019art,zeller2019fuzzing,bohme2020boosting,ba2022stateful,shah2022mc2,zhu2022fuzzing,qian2023dipri} uses lightweight program instrumentation to track test input propagation through the control flow graph (CFG). Greybox fuzzers such as \texttt{AFL++} \cite{fioraldi2020afl++}, \texttt{libFuzzer} \cite{libfuzzer}, and \texttt{FuzzTest} \cite{fuzztest} insert ``trampolines'' at compile time after conditional jumps that identify branches and (approximately) increment associated counters. They then operate as in Figure \ref{fig:greyboxFuzzingArchitecture}. 

This ``go-explore'' loop both anticipated, and can benefit from, recent developments in quality-diversity optimization algorithms \cite{ecoffet2021first,huntsman2023quality}. It is also informed by the CFG geometry as captured by, e.g. a remarkable metric on a Markov chain that models the joint dynamics of a fuzzer and program \cite{bohme2017directed,boyd2021metric}: see Figures \ref{fig:MarkovMetric} and \ref{fig:MarkovMetricDetail}.

A greybox fuzzer's goal is to reach certain parts of the CFG: coverage-based greybox fuzzing is merely a uniform instance of such directed greybox fuzzing (DGF) \cite{bohme2017directed}. Viewing this goal as intrinsically geometrical complements recent mathematical abstractions of DGF as a probabilistic or information-theoretical process \cite{bohme2017directed,bohme2020boosting,shah2022mc2}. In particular, the fuzzer tries to find diverse inputs and paths. This presupposes a heretofore often implicit notion of dissimilarity both on the input space and the CFG itself; it also suggests viewing fuzzing more explicitly as an optimization problem.

We take these ideas seriously and literally by prototyping and evaluating a prototype fuzzer called \texttt{GoExploreFuzz} \cite{goExploreFuzz}. It leverages the mathematics of diversity that can be characterized as geometrical information theory \cite{posada2020gait,leinster2021entropy}. However, because the fundamental data structures and algorithms in this vein respectively require quadratic and cubic time (as a function of input corpus size) to populate and execute in practice, this perspective is only likely to be immediately fruitful for programs whose execution is time-intensive, say, on the order of seconds to minutes. Nevertheless, such situations are still of practical interest and import and can arise, e.g., because uninstrumented function calls require time to execute, or when environmental inputs \cite{ba2022stateful} require time to acquire.

\texttt{GoExploreFuzz} borrows several algorithmic constructions from the breakthrough Go-Explore algorithm of \cite{ecoffet2021first} as formalized in \cite{huntsman2023quality}. In the process of exploring the underlying ideas, we find that an important impact of geometry on DGF is the ability to autonomously tune the scale or \emph{bandwidth}\footnote{
We borrow this term from kernel density estimation in statistics, where it roughly corresponds to the standard deviation of (say) a Gaussian.
}
of mutations based on the geometry of input space as discretized by ranked proximity to a maximally diverse subset of CFG paths (see Figure \ref{fig:cells}). We also find that optimizing the objective of \emph{hitting probability distance} \cite{boyd2021metric} from the program entry point yields good coverage. Taken together, these and associated findings indicate promising directions for fuzzing programs whose execution is time-intensive.

\section{\label{sec:fuzzingGeometry}Geometry of greybox fuzzing}

A geometrical perspective on greybox fuzzing would be aided by a dissimilarity on vertices of a CFG that incorporates the history of a given fuzz campaign. Such a dissimilarity should properly only deal with the CFG structure and the statistics of exercised paths, not white-box information such as details of basic blocks. The attendant geometry can in turn determine the current queue or distribution of promising inputs maintained by the fuzzer's curator. To promote coverage, the curator can preferentially select outliers, while to reach a particular edge or vertex, the curator can preferentially select promising inputs whose corresponding paths are closer.

More generally, geometrical considerations should properly inform each of  four key aspects of greybox fuzzing, viz. 
\begin{itemize}
	\item queue scheduling or input sampling (see just above);
	\item power schedules, i.e., effort spent mutating an input;\footnote{Our experiments indicate that this aspect actually has the least evidence supporting the use of a \emph{nontrivial} notion of geometry. That is, the so-called \emph{discrete metric} $d(x,x') = 1$ if $x \ne x'$ appears to be generically appropriate, as manifested in the use of the \texttt{ENTROPIC} power schedule versus a (dis)similarity-sensitive generalization.
	}
	\item mutation operators: mutants should be neither too similar to nor too dissimilar from a generating input, and this quantification is necessarily dependent on both the given input and the corpus it is part of;
	\item assessement of input behaviors: unusual behaviors correspond to outliers in some reasonable notion of geometry.  
\end{itemize}
Recently developed ideas linking diversity, geometry, and information theory under the guise of \emph{magnitude} \cite{leinster2021entropy} allow us to address all four of these aspects given a reasonable symmetric dissimilarity. There are two technical obstacles to this: first, CFGs are manifestly not symmetric, and second, the underlying algorithms are compute-intensive, relying heavily on dense linear algebra. However, we will demonstrate below that the hitting probability distance of \cite{boyd2021metric} solves the first of these problems (see Figures \ref{fig:MarkovMetric} and \ref{fig:MarkovMetricDetail}). Meanwhile, restricting consideration to target programs whose execution is time-intensive renders the second problem irrelevant.

The vehicle for an initial realization of these ideas is our prototype fuzzer \texttt{GoExploreFuzz} \cite{goExploreFuzz}. It exploits the similarity between greybox fuzzing and the framework of \cite{ecoffet2021first}, viz.
\begin{itemize}
	\item select and {\bf go} to a promising or \emph{elite} input;
	\item {\bf explore} a neighborhood of the selected elite;
	\item map resulting inputs to a discretization of space into \emph{cells};
	\item update the elites in populated cells according to the values of an objective function to be optimized.
\end{itemize}
The theory of magnitude gives a principled foundation for addressing the first three of these tasks, and an implementation of Go-Explore built on this foundation already enables efficient quality-diversity optimization of objective functions whose execution is time-intensive \cite{huntsman2023quality}. However, an application to greybox fuzzing \emph{per se} requires several major modifications.

\section{\label{sec:goExploreFuzzDesign}\texttt{GoExploreFuzz}}

The Go-Explore implementation of \cite{huntsman2023quality} is designed for optimizing objectives using an efficient approximation. The code itself relies on radial basis function interpolation, 
though any other 
approximation technique can substitute. There is, however, a deeper obstacle to adapting the work of \cite{huntsman2023quality} for fuzzing. While \cite{huntsman2023quality} exhibits toy examples of fuzzing using the geometry of a CFG induced by a graph drawing or distance to a particular vertex, these examples all fail to address the fundamental issue that once a program's behavior can be efficiently approximated on a local set of inputs, additional fuzzing there is counterproductive \cite{nicolae2023revisiting}.

\begin{figure}[h]
  \centering
    \begin{tikzpicture}[
      every node/.style={on grid},
      setA/.style={fill=black,circle,inner sep=2pt},
      setC/.style={fill=black,rectangle,inner sep=2pt},
      every fit/.style={draw,fill=gray!15,ellipse,text width=25pt},
      >=latex
    ]
    
    \node[setA,label=left:$x_1$] (a) {};
    \node [setA,below = of a,label=left:$x_2$] (b) {};
    \node [setA,below = of b,label=left:$x_3$] (c) {};
    \node[above=of a,anchor=south] {inputs $\in X$};
    
    \node[inner sep=0pt,right=3cm of a] (x) {$y_1$};
    \node[below = of x] (y) {$y_2$};
    \node[inner sep=0pt,below = of y] (z) {$y_3$};
    \node[above=of x,anchor=south] {paths $\in Y$};
    
    \node[setC,label=right:$z_1$,right = 3cm of x] (m) {};
    \node[setC,label=right:$z_2$,below = of m] (n) {};
    \node[setC,label=right:$z_3$,below = of n] (p) {};
    \node[above=of m,anchor=south] {objective values $\in \mathbb{R}$};
    
    \draw[->,shorten >= 3pt] (a) -- node[label=above:$\gamma$] {} (x);
    \draw[->,shorten >= 3pt] (b) -- node[label=above:$\gamma$] {} (x);
    \draw[->] (c) -- node[label=above:$\gamma$] {} (y);
    \draw[->,shorten <= 3pt] (x) -- node[label=above:$\phi$] {} (p);
    \draw[->] (y) -- node[label=above:$$] {} (n);
    \draw[dashed,->] (z) -- node[label=above:$$] {} (m);
    
    \begin{pgfonlayer}{background}
    \node[fit= (a)  (c) ] {};
    \node[fit= (x) (z) ] {};
    \node[fit= (m) (p)] {};
    \end{pgfonlayer}
    \end{tikzpicture}
\caption{Directed greybox fuzzing sends inputs to paths to objective values. An unrealized path is shown with dashes.
  }
  \label{fig:directedGreyboxFuzzing}
\end{figure}
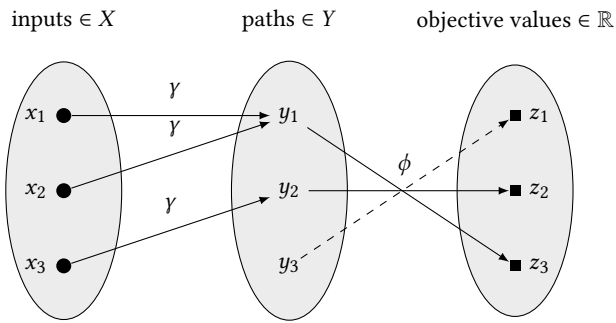

In short, directed greybox fuzzing has special structure as illustrated in Figure \ref{fig:directedGreyboxFuzzing} that \texttt{GoExploreFuzz} respects by i) avoiding a surrogate/metamodel for the objective and ii) adding an intermediate ``phenotype'' space. The fuzzer acts as a composition of two functions $\gamma : X \rightarrow Y$ (``genotype'') and $\phi : Y \rightarrow \mathbb{R}$ (``phenotype'' or ``fitness'') that respectively perform dynamic execution and objective evaluation (e.g., proximity to a set of desired locations, or a constant for the case of pure coverage). Here the sets of inputs $X$ and CFG paths $Y$ are respectively endowed with dissimilarity measures $d_X$ and $d_Y$. We assume that $\gamma$ takes much longer than $\phi$ to run and that $d_X$ is fast (e.g., a Hamming distance). 

In order to apply the theory of magnitude towards maximizing diversity, we require that $d_X$ and $d_Y$ be symmetric \cite{leinster2016maximizing,leinster2021entropy}. $d_X$, which we assume is fast to run, is mainly useful for tuning the bandwidth of mutations, which is often conveniently characterized as a number of sequentially applied primitive or atomic mutations. 

\begin{figure}[h]
  \centering
  \includegraphics[trim = 15mm 110mm 15mm 110mm, clip, width=\columnwidth,keepaspectratio]{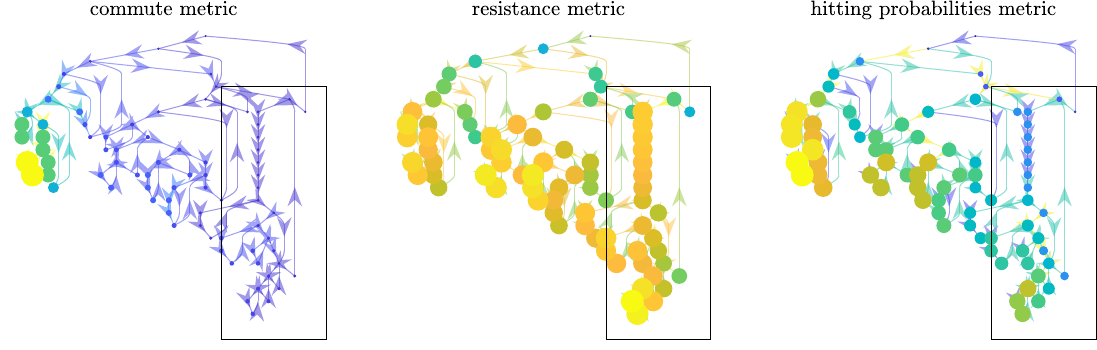}
\caption{(Left) Commute time metric on the Markov chain corresponding to a uniform random walk on the (strongly connected) CFG shown. Edges are colored from blue (nearest) to yellow (farthest) according to distance from source to target; vertex sizes and colors indicate distance from the program entry (at line 1 and the top of the drawings). The rectangle indicates the scope of detail for Figure \ref{fig:MarkovMetricDetail}. (Center; right) As in the left panel, but for resistance and hitting probability metrics. More detail is shown in Figure \ref{fig:MarkovMetricDetail}.
  }
  \label{fig:MarkovMetric}
\end{figure}

\begin{figure}[h]
  \centering
  \includegraphics[trim = 10mm 90mm 10mm 90mm, clip, width=.968\columnwidth,keepaspectratio]{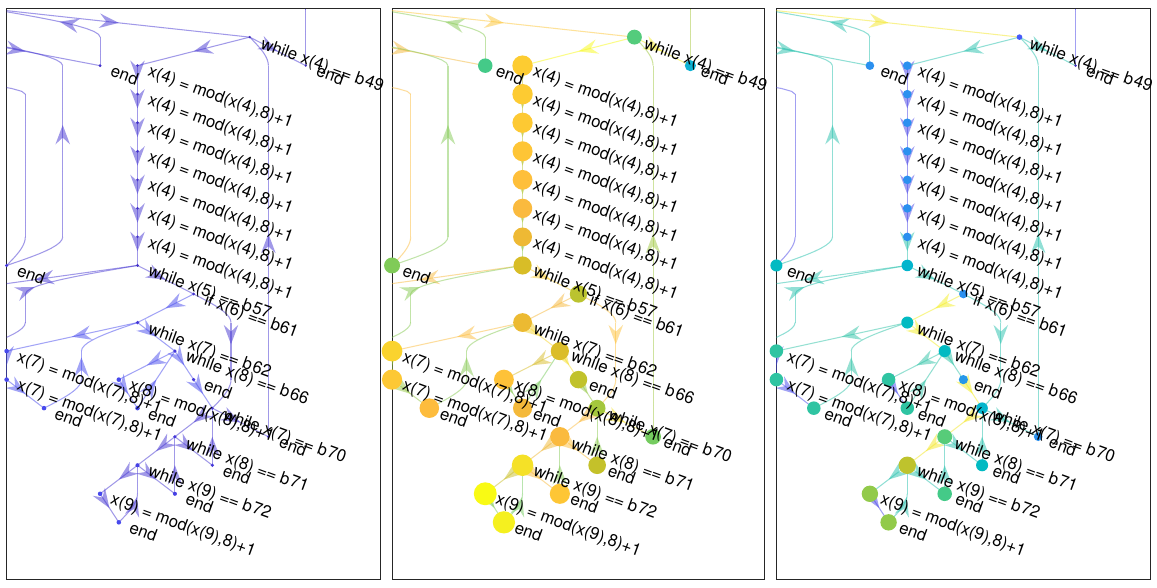}
\caption{Details of corresponding panels in Figure \ref{fig:MarkovMetric}. The commute (respectively, resistance) metrics have bad behavior in the lower (respectively, upper) area, while the hitting probability metric behaves well everywhere. The actual program instructions account for the indexing of inputs and nesting of predicates via a depth-first-search of the CFG; assignments amount to cyclic increments of words that (along with \texttt{b} variables) can take values in $\{1,\dots,N\}$: here, $N = 8$.
  }
  \label{fig:MarkovMetricDetail}
\end{figure}

Meanwhile, $d_Y$ is used to promote path diversity, which is \emph{a priori} more directly relevant for coverage than diversity with respect to $d_X$ as implicitly considered in \cite{qian2023dipri}. We can afford to spend considerable effort evaluating $d_Y$ towards this end because $\gamma$ is time-intensive anyway. Although a CFG is  asymmetric, the recently developed hitting probability metric of \cite{boyd2021metric} on a Markov chain is symmetric and well-behaved, allowing us to work with the empirical random walk on the CFG produced by the fuzzer so far. While there are other symmetric dissimilarities and \emph{bona fide} metrics available, e.g., commute times or resistance metrics \cite{young2015new}, their phenomenology is undesirable for our purposes: see Figures \ref{fig:MarkovMetric} and \ref{fig:MarkovMetricDetail}.

With a phenomenologically satisfactory metric on a CFG in hand, we can lift it to paths in several ways that allow us to quantify diversity contributions of paths. One way is to use the Hausdorff distance (i.e., the usual notion of distance between subsets of a metric space) on sets of vertices along paths: this is simple and comparatively cheap to run. Another way is to use an edit distance on paths: this is elegant (though it requires one slightly arbitrary choice of insertion/deletion cost) but expensive to run. Many other lifts are possible as well. If as in our case the CFG metric evolves over time, the diversity contributions of paths will as well. As a fuzz campaign progresses, the salience of the lifted geometry increases.

\subsection{\label{sec:details}Details}

\begin{figure}[h]
  \centering
  \includegraphics[trim = 80mm 110mm 77mm 109mm, clip, width=.49\columnwidth,keepaspectratio]{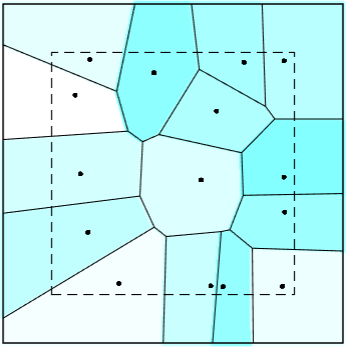}
  \includegraphics[trim = 80mm 110mm 77mm 109mm, clip, width=.49\columnwidth,keepaspectratio]{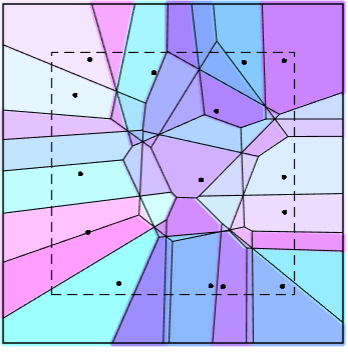}
\caption{(Left) Using Euclidean distance for illustration purposes, an online algorithm of \cite{huntsman2023quality} selects 15 landmarks (black dots) from 41 candidates that are sampled uniformly at random in the dashed square. Note that these landmarks are mostly near the periphery of the square. Each shaded region corresponds to a cell defined by the set of points with given nearest landmark. The geometry of an elite input in a cell determines the bandwidth for mutations. (Right) Each shaded region corresponds to a cell defined by the set of points with given nearest and second nearest landmarks.
  }
  \label{fig:cells}
\end{figure}

\texttt{GoExploreFuzz} \cite{goExploreFuzz} takes as arguments a set of initial inputs and a subset thereof whose corresponding paths are diverse \emph{landmarks} that nicely discretize path space into \emph{cells} by ranking the dissimilarities of a path to each of the landmarks as schematically indicated in Figure \ref{fig:cells}. While \cite{huntsman2023quality} handles this internally as part of its initialization, obtaining diverse paths in the CFG may require many more program executions than in the pure quality-diversity/optimization context. Also, it may be necessary to use the results of one fuzz campaign to start a subsequent one, so we separate these concerns.\footnote{
Because program execution $\gamma$ is presumably expensive, \texttt{GoExploreFuzz} also takes $\gamma(\text{initial corpus})$ as a precomputed input.
}

Other inputs include the execution map $\gamma : X \rightarrow Y$, the objective $\phi: Y \rightarrow \mathbb{R}$, respective dissimilarities $d_X$ and $d_Y$ on inputs and paths, and anonymous functions for updating data on $Y$ (e.g., CFG edge counts), producing initial unmutated inputs, and mutating inputs. Finally, \texttt{GoExploreFuzz} takes a predetermined budget for the number of overall program executions/evaluations to perform (because they are time-intensive) and a bound on the power schedule.

The power schedule itself defaults to the exploration effort of \cite{huntsman2023quality}, which is akin to the \texttt{AFL++} default. However, \texttt{GoExploreFuzz} also has options for the state-of-the-art \texttt{ENTROPIC} power schedule \cite{bohme2020boosting} as well as a (dis)similarity-sensitive variant called \texttt{SIMTROPIC}. 

The curator uses probabilistic sampling from a ``go distribution'' rather than a queue. Besides the default of \cite{huntsman2023quality} that balances diversity/exploration and quality/exploitation, there are seven other variants (B-H in the legends of Figures \ref{fig:averageOverCFGs20230809}-\ref{fig:averageOverCFGsAtEvalBudget2D20230809}). As we shall see, these have little impact apart from a pure exploration mechanism that performs badly, so we omit further details here: for code, see \cite{goExploreFuzz}.

Two additional options are important to mention. One (K in the legends of Figures \ref{fig:averageOverCFGs20230809}-\ref{fig:averageOverCFGsAtEvalBudget2D20230809}) turns off the bandwidth adaptation mechanism. The other (L in the legends of Figures \ref{fig:averageOverCFGs20230809}-\ref{fig:averageOverCFGsAtEvalBudget2D20230809}) avoids downselecting mutants using biobjective Pareto domination of diversity measures relative to elites and other mutants.

\section{\label{sec:experiment}Experimental evaluation}

Because \texttt{GoExploreFuzz} is a preliminary architectural proof of concept developed in \texttt{MATLAB} for fuzzing time-intensive programs that would be intrinsically resistant to statistical evaluation, we did not attempt to benchmark it. Instead, we used various configurations and objectives on a suite of toy programs that allowed us to draw robust conclusions about ideas in the context of our framework.

\begin{figure}[h]
  \centering
\includegraphics[trim = 24mm 72mm 24mm 72mm, clip, width=\columnwidth,keepaspectratio]{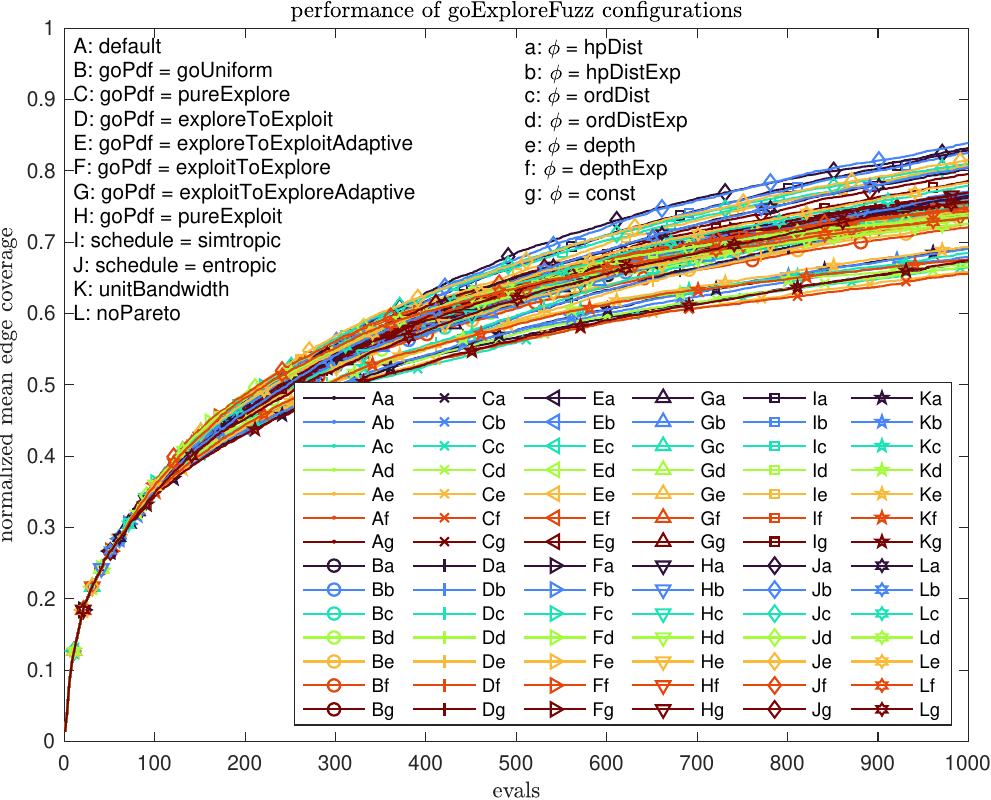}
\caption{Normalized mean edge coverage vs. program executions/evaluations under configurations A-L and objectives a-g. Standard deviations at the end (not shown) are all close to 0.1, and behavior of configuration/objective pairs across various programs is relatively consistent (also not shown).
  }
  \label{fig:averageOverCFGs20230809}
\end{figure}

\begin{figure}[h]
  \centering
\includegraphics[trim = 30mm 90mm 60mm 90mm, clip, width=.75\columnwidth,keepaspectratio]{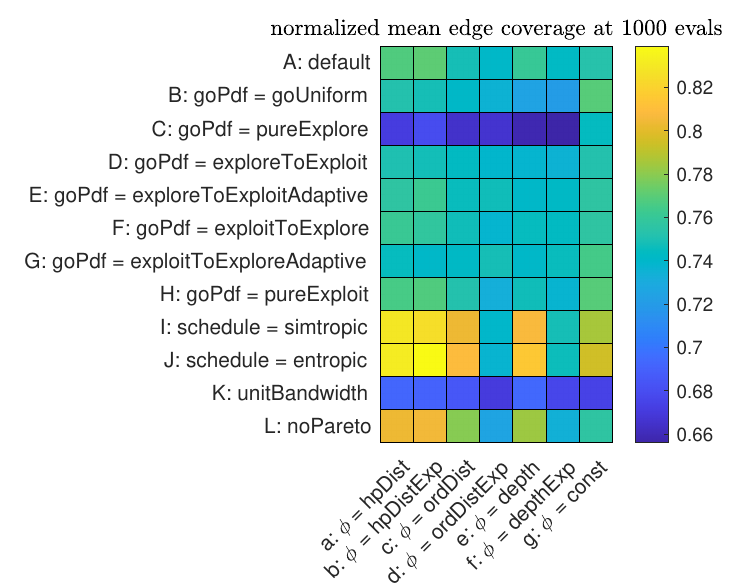}
\caption{Normalized mean edge coverage at the total program execution/evaluation budget of 1000.
  }
  \label{fig:averageOverCFGsAtEvalBudget2D20230809}
\end{figure}

Figures \ref{fig:averageOverCFGs20230809}-\ref{fig:averageOverCFGsAtEvalBudget2D20230809} show the results of a comprehensive test of programs, configurations and objectives as detailed in \cite{goExploreFuzz}. We considered 10 toy programs, ran 40 fuzz campaigns per program, with 1000 program evaluations per campaign, used 12 different configurations (A-L in legends) and seven different objectives (a-g in legends: all but the constant objective indicated by g are different notions of distance from the program entry or depth in a drawing of the CFG). {\bf All told, this test involved 33.6 million program evaluations, requiring over 20 calendar days using all eight Intel Core i9 parallel cores on a macOS system with 64 GB of RAM.}

We produced the tested programs from a stochastic context free grammar replacing nonterminals with \texttt{if}, \texttt{while}, or assignment constructs. We also implemented a toy compiler front end that formed control flow graphs. Finally, we constructed assignments and predicates in a way that guaranteed no code was unreachable. The structure of the assignments and predicates generalizes the toy example in \cite{bohme2016coverage}: in particular, achieving coverage amounts to guessing ``good'' entries of inputs in sequence. Because \texttt{MATLAB} is an interpreted language, we simply dynamically executed the programs using \texttt{eval}, with inline greybox instrumentation for traces.

Inspection of Figures \ref{fig:averageOverCFGs20230809}-\ref{fig:averageOverCFGsAtEvalBudget2D20230809} reveals several clear conclusions:
\begin{itemize}
	\item[A:] the default ``go distribution'' balancing exploration and exploitation seems to work at least as well as any other;
	\item[J:] the \texttt{ENTROPIC} power schedule clearly improves upon the default, but the (dis)similarity-sensitive analogue \texttt{SIMTROPIC} provides no additional advantage;
	\item[K:] tuning mutation bandwidth is very useful;
	\item[L:] it is harmful to use mutants that are only diverse with respect to elites or their fellow mutants, as it inhibits exploitation of inputs that are ``almost right'';
	\item[a:] the objective of hitting probability distance from a program entry yields good coverage that is slightly more robust if otherwise slightly worse than its exponential (b); these improve on ``ordinary'' digraph distance and depth in a CFG drawing (and their exponentials and a constant).
\end{itemize}

The failure of a (dis)similarity-sensitive version of entropy (viz., diversity) to improve on \texttt{ENTROPIC} is unsurprising in hindsight. Unpacking the rationale behind \texttt{ENTROPIC}, the underlying geometry turns out to be discrete/trivial. Consequently, the diversity (of order 1) reduces to ordinary Shannon entropy.

\section{\label{sec:conclusion}Conclusion}

By discretizing the spaces of inputs and paths, a directed greybox fuzzer can achieve broad coverage using a generic objective such as hitting probability distance from program entry. Our evaluation is fairly general since a CFG can be restructured using the algorithm of \cite{zhang2004using} to have (planar structure and) semantics like our test programs, and we can pull back an objective to the original CFG or work with the restructured CFG. Finally, a geometrical perspective on directed greybox fuzzing is likely to be useful for time-intensive programs.

\begin{acks}
Thanks to Megan Fuller, Zac Hoffman, Rachelle Horwitz-Martin, Matt Kornbluth, and Mike Richman for discussions and suggestions. This research was developed with funding from the Defense Advanced Research Projects Agency (DARPA). The views, opinions and/or findings expressed are those of the author and should not be interpreted as representing the official views or policies of the Department of Defense or the U.S. Government. Distribution Statement “A” (Approved for Public Release, Distribution Unlimited). 
\end{acks}

%
\bibliographystyle{ACM-Reference-Format}

\end{document}